\documentclass[12pt]{article}
\usepackage{graphicx} 

\def\abstract#1{{\vspace*{0.9 cm}
    \centering{\begin{minipage}{12.2truecm}\footnotesize\baselineskip=12pt\noindent
        \centerline{\footnotesize ABSTRACT}\vspace*{0.3cm}
        \parindent=0pt #1
      \end{minipage}}\par}} 

\title{Topological Study of Three-Jet Events in ALICE}
\author{Sona Pochybova\ for the ALICE collaboration\\
         MTA KFKI RMKI\\
         29-33 Konkoly-Thege Miklos Str., Budapest, Hungary\\
         e-mail: sona.pochybova@cern.ch}
\begin{document}
\maketitle
\abstract{
The ALICE experiment at LHC is dedicated to study matter formed in heavy-ion collisions, but also has a strong physics program for $pp$ collisions. In these collisions, protons will collide at energies never reached before under laboratory conditions. At the high energies, ALICE will enable us to study jet physics in detail, especially the production of multiple jet events, setting the baseline for heavy-ion. Three-jet events allow us to examine the properties of quark and gluon jets, providing a suitable tool for testing QCD experimentally.
We discuss the selection method and topology of three-jet events in ALICE. The analysis was performed on two PYTHIA data sets, both involving $pp$ collisions at $\sqrt{s} = 14$ TeV with enhanced jet production. The results from the dedicated jet MC production are discussed and compared to previous studies at CDF and D\O. We investigate the possibilities to determine gluon jet candidates.
}
\section{Introduction}
\label{sec:Introduction}
In perturbative QCD, constituent partons in $pp$ collisions experience hard scatterings. Outgoing scattered partons branch via soft quark and gluon radiation and this way, they form showers. This process is called fragmentation. The showers eventually hadronise and can be observed experimentally as colourless hadrons inside jet cones. If during the stage of fragmentation one of the partons radiates a hard gluon, one can expect to see a three-jet event.
\paragraph{}Through proper reconstruction of such events, a possibility emerges to study the properties of initial partons and the conditions they evolved in. Furthermore, through the identification of original partons, one can examine fragmentation properties of quark and gluon jets separately. Three jet events with the variety of processes ($gg\rightarrow ggg$, $q\bar{q}\rightarrow ggg$, $q\bar{q}\rightarrow q\bar{q}g$ and those related by crossing symmetry) and topologies involved are suitable for thorough investigation and testing of higher order QCD. 
\paragraph{}At LHC, protons will collide at energies never reached before and thus, opening doors to new physics and further scrutiny of our present knowledge. In these proceedings we present preliminary results on topological studies of three-jet events in $pp$ collisions at $\sqrt{s}=14$ TeV and discuss a probabilistic approach to determine quark and gluon jet candidates.
\section{Quark and gluon jet differences}
Differences between quark and gluon initiated jets stem from the partons different colour factors, sometimes referred to as colour charges. Gluon carries colour factor ${\textrm C_{A}} = 3$ and quark carries colour factor ${\textrm C_{F}} = 4/3$, hence:
\begin{equation}
\label{eq:ColorFactors}
\frac{{\textrm C_{A}}}{{\textrm C_{F}}} = \frac{9}{4}
\end{equation}
The colour factors are proportional to the probability that a parton radiates a soft gluon. Colour factor of a gluon is $9/4$ times bigger than the one of a quark in asymptotic limit of $Q^{2}\rightarrow \infty$. This means, that gluons branch more easily and form higher multiplicity jets. As a consequence of the higher multiplicity, gluon jets are expected to be broader with softer fragmentation function than quark jets. All these differences can be expressed in terms of $C_{A}/C_{F}$ ratio (Eq.~\ref{eq:ColorFactors})\  \cite{Scripta:Weber, Ellis:1991qj}.
%
%
\paragraph{} Quark and gluon jet properties were previously studied by many collaborations in various colliding systems; $ep$, $e^{+}e^{-}$ and $p\bar{p}$. Most of the studies were conducted at LEP $e^{+}e^{-}$ collider, for example see Refs.~\cite{DELPHI:1996,Barate:1998cp}. The source of gluon jets in these experiments were $Z^{0}\rightarrow q\bar{q}$ decays in which a hard gluon was radiated by either quark or anti-quark. To identify gluons in such events, the method of $b$-tagging was widely used in two and three fold symmetry events; "Y" and "Mercedes" types respectively. The OPAL collaboration also performed a study with light quarks \cite{Alexander:1995bk}.
\paragraph{}In $e^{+}e^{-}$ collisions, as mentioned above, the only source of gluon jets are specific types of events. Hadron collisions, on the other hand, involve a wider range of jet production processes (see Tab.~$\ref{tab:PythiaProc}$) and as such, provide a rich source of gluon jets. Studies conducted by CDF and D\O~ collaborations \cite{Affolder:2001jx,Abazov:2001yp}, in $p\bar{p}$ collisions at $\sqrt{s} = 1.8$ TeV, examined quark and gluon jet properties in di-jet events. Gluon jet candidates were chosen on statistical basis and compared to quark jets from $\gamma -q$ and $Z-q$ events.   
\paragraph{}Experimental data collected in previous years, eg. by LEP and Tevatron collaborations, indeed show different multiplicities and fragmentation properties of quark and gluon initiated jets. Nevertheless, as the energy scales achieved were below the asymptotic limit, quantitatively these differences did not reach the predictions given by the colour factor ratio (Eq.~\ref{eq:ColorFactors}). Study of quark and gluon jets in three jet events at LHC represents a further step in reaching the expected values as well as probing them at higher orders.
\section{Analysis}
\subsection{Topology of three-jet events}
\begin{figure}[ht]
\centering
\includegraphics{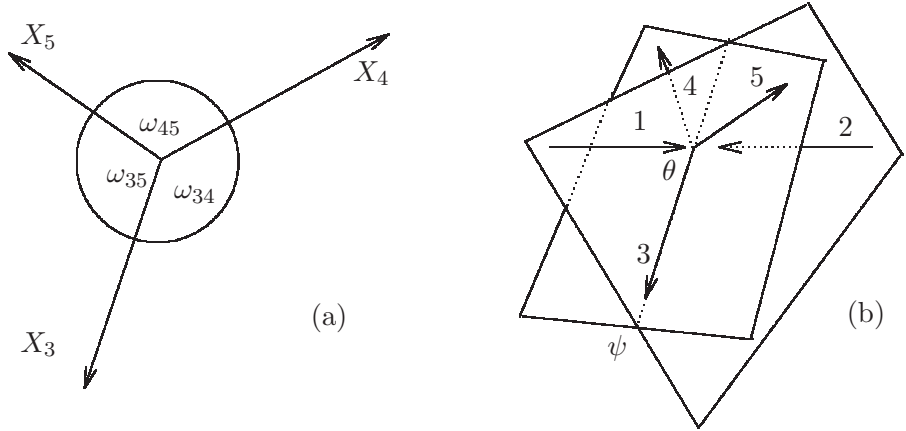}
\caption{Illustration of the topological variables. Panel (a): Dalitz variables and angles between jets, panel (b): angles with respect to incoming partons $1$ and $2$.}
\label{fig:TopVar}
\end{figure}

\begin{figure}[ht]
\centering
\includegraphics[width=100mm,height=90mm]{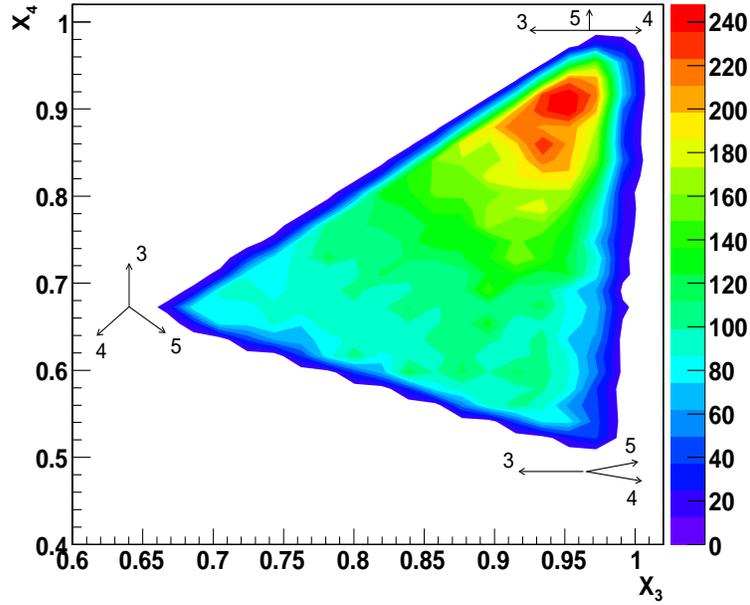}
\caption{Dalitz plane for three-jet events found using PYCELL routine inside PYTHIA (see text for details). In each corner of the triangle a corresponding topological configuration is drawn.}
\label{fig:PYTHIADalitz}
\end{figure}

The topology of a final three parton system in hadron collisions can be described using eight variables. All of these are calculated in the rest frame of the final state three parton system. Incoming partons are labelled $1$ and $2$ with energies ${\textrm E_{1}}$ and ${\textrm E_{2}}$ respectively (${\textrm E_{1}} > {\textrm E_{2}}$). Outgoing partons are labelled $3$, $4$ and $5$ with energies ${\textrm E_{3}}$, ${\textrm E_{4}}$ and ${\textrm E_{5}}$ (${\textrm E_{3}} > {\textrm E_{4}} > {\textrm E_{5}}$), where parton 3 is also referred to as the leading parton. The partons are associated with momentum vectors $\vec{\textrm p_{i}},\ i=1,..5$. Topological variables can be defined as follows \cite{Abachi:1995zv, Acosta:2004yy}:
\begin{itemize}
\item Cosine of the angle between the beam and the leading parton:
\begin{equation}
\label{eq:CosineTheta3Def}
        \cos{\theta} = \frac{\vec{p_{1}}.\vec{p_{3}}}{|\vec{p_{1}}||\vec{p_{3}}|},
\end{equation}
\item Cosine of the angle between plane containing the leading parton $3$ and the incoming parton $1$, and the plane containing partons $4$ and $5$:
\begin{equation}
\label{eq:CosinePsiDef}
\cos{\psi} = \frac{\left(\vec{p_{1}}\times \vec{p_{3}}\right).\left(\vec{p_{4}}\times \vec{p_{5}}\right)}{|\vec{p_{1}}\times \vec{p_{3}}|.|\vec{p_{4}}\times \vec{p_{5}}|}
\end{equation}

\item Fractions of energies carried by each of the final state partons, called the {\it Dalitz variables}:
\begin{equation}
\label{eq:DalitzDef}
    X_{i} = \frac{2\cdot E_{i}}{\sum_{i}E_{i}},~i=3, 4, 5
\end{equation}
\item  Fractions of energies carried by a pair of partons depending on the opening angle between them, so called {\it scaled masses}:
\begin{equation}
\label{eq:ScaledMassDef}
        \mu_{ij} = \frac{m_{ij}}{\sqrt{\hat{s}}} = \sqrt{X_{i}X_{j}\left(1-\cos{\omega_{ij}}\right)/2},
\end{equation}
where $\mu_{ij}$ ($i,j=3,4,5$) are the invariant masses of final state parton pairs, $\omega_{ij}$ ($i,j=3,4,5$) are the angles between partons, and $\sqrt{\hat{\textrm s}}$ is the invariant mass of the final state three parton system.
\end{itemize}
The two angles (Eq.~\ref{eq:CosineTheta3Def} and \ref{eq:CosinePsiDef}) determine the relative position of the final parton system with respect to the incoming partons (Fig.~\ref{fig:TopVar}, panel (b)). The energy fractions (Eq.~\ref{eq:DalitzDef} and \ref{eq:ScaledMassDef}), together with angles between partons, provide information on the event shape (Fig.~\ref{fig:TopVar}, panel (a)). Experimentally, these variables are calculated for jets found by jet-finding algorithms.
Our study is not concerned with all of the variables listed above, merely we focus on Dalitz variables ($X_{3}$, $X_{4}$) and scaled masses ($\mu_{ij},\ i,j=3,4,5$).
\paragraph{}
The topology of three jet events was previously studied at Tevatron by D\O ~and CDF collaborations \cite{Abachi:1995zv, Acosta:2004yy}. Three-jet events found in these experiments were distributed over the Dalitz plane with an enhancement in $X_{3}\rightarrow 1$ and $X_{4}\rightarrow 1$. With these values of Dalitz variables, the event shape will look like as shown in Fig.~\ref{fig:PYTHIADalitz}, in the $[X_{3}, X_{4}] = [1,1]$ corner.
Topological attributes of three-jet events in these studies agree with theoretical predictions.
\subsection{Sample and event selection}
Events were generated using PYTHIA $6.4$ generator \cite{PYTHIA:MANUAL} included in the AliRoot framework. Samples consist of $pp$ collisions at $\sqrt{s} = 14$ TeV with CTEQ5L parton distribution functions. Events were simulated in two ${\textrm p_{t}^{hard}}$ bins, namely ${\textrm p_{t}^{hard}} > 50$ GeV/c and ${\textrm p_{t}^{hard}} > 100$ GeV/c, where ${\textrm p_{t}^{hard}}$ is the transverse momentum of partons in the rest frame of the hard scattering \cite{ALICE:PPR2}.
\paragraph{}
To reconstruct jets, a UA1 cone based algorithm is applied to the charged particles after reconstruction. More details on this jet finder can be found in \cite{ALICE:PPR2} and \cite{Blazey:2000qt}. The input parameters used for jet reconstruction are summarised in Tab.~\ref{tab:ConeParams}. If three jets (each with energy ${\textrm E_{t}} > 5$ GeV) are found within the fiducial region $|\eta| < 0.5$, then such event is accepted. From $142000$ studied events, $3080$ were found to fulfil these conditions.
\begin{table}[ht]
\begin{minipage}[b]{0.48\linewidth}
\centering
\begin{tabular}{lc}\hline
Parameter &Value\\\hline\hline
{\it R}&$0.5$\\
{\it E(seed)}&$2.0$~GeV\\
$E_{t_{min}}$&$5.0$~GeV\\
$p_{t_{min}}$&$1.5$~GeV/c\\\hline
\end{tabular}
\caption{Values of input parameters used for analysis. $R$ is the cone size, $E(seed)$ is the energy of the seed, $E_{t_{min}}$ is the minimal energy of the reconstructed jet and $p_{t_{min}}$ stands for minimal transverse momentum of the tracks inside the jet.}
\label{tab:ConeParams}
\end{minipage}
\hspace{0.5cm}
\begin{minipage}[b]{0.48\linewidth}
\centering
\begin{tabular}{llll}\hline
ID&Processes&ID&Process\\\hline\hline\vspace{0.2cm}
$11$&$q_{i}q_{j}\rightarrow q_{i}q_{j}$&$28$&$q_{i}g\rightarrow q_{i}g$\\\vspace{0.2cm}
$12$&$q_{i}\bar{q_{i}} \rightarrow q_{k}\bar{q_{k}}$&$53$&$gg\rightarrow q_{k}\bar{q_{k}}$\\\vspace{0.2cm}
$13$&$q_{i}\bar{q_{i}}\rightarrow gg$&$68$&$gg\rightarrow gg$\\\hline
\end{tabular}
\caption{$2\rightarrow 2$ processes used in PYTHIA to generate QCD jets. The $2\rightarrow 3$ processes are generated adding ISR (Initial State Radiation) or FSR (Final State Radiation) \cite{PYTHIA:MANUAL}.\newline}
\label{tab:PythiaProc}
\end{minipage}
\end{table}
\paragraph{}The UA1 finder does not implement splitting and merging of the found jets, therefore, a particle may be assigned to more than one jet. To avoid this a separation cut is introduced. Separation of jets is defined as the distance between jet axes in $\eta-\phi$ plane; $\Delta R = \sqrt{\left(\Delta \eta\right)^{2}+\left(\Delta \phi\right)^{2}}$. All the events containing pairs of jets with separation smaller than twice the cone size, {\it i.e} $\Delta R_{ij} < 1.0$, were filtered out. After the cut, the remaining sample included 1108 reconstructed events. 
\paragraph{}The data presented are raw, without energy corrections. We compare the reconstructed jets to generated and PYTHIA jets. The generated ones represent jets found by applying UA1 cone based algorithm to final state particles after generation. Same settings apply as for the reconstructed jets. PYTHIA jets refer to jets found by PYCELL routine embedded inside PYTHIA. This routine is a cone based jet finder, that is applied to the final state particles after generation \cite{PYTHIA:MANUAL}. The cone size used is $R=1$ and jets within $|\eta| < 2.0$ are accepted.   
\subsection{Results and discussion}
\begin{figure}[ht]
\centering
\includegraphics[width=150mm,height=75mm]{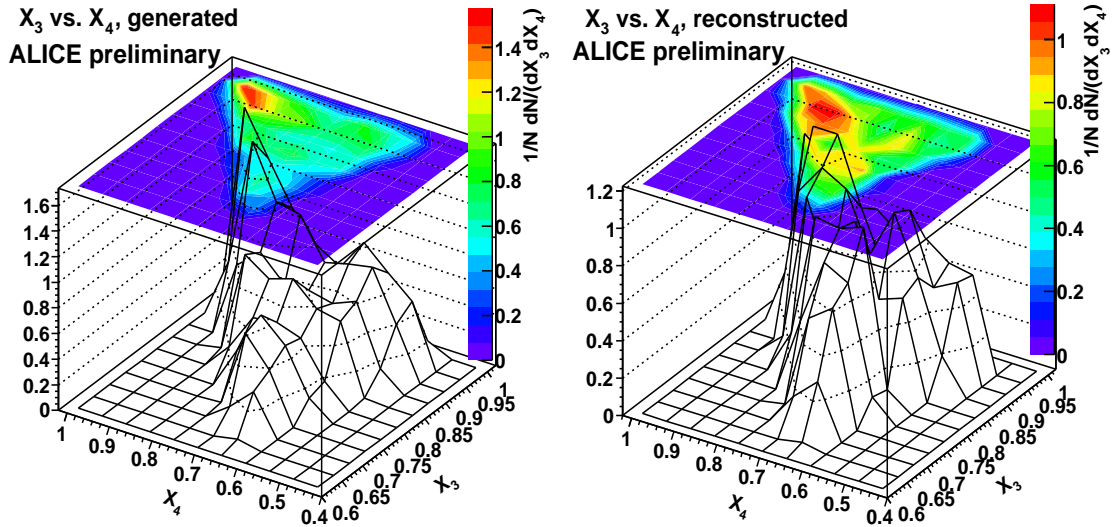}
\caption{Dalitz plane for generated (left) and reconstructed (right) three-jet events.}
\label{fig:ALICEDalitz}
\end{figure}

\begin{figure}[ht]
\centering
\includegraphics[width=150mm,height=50mm]{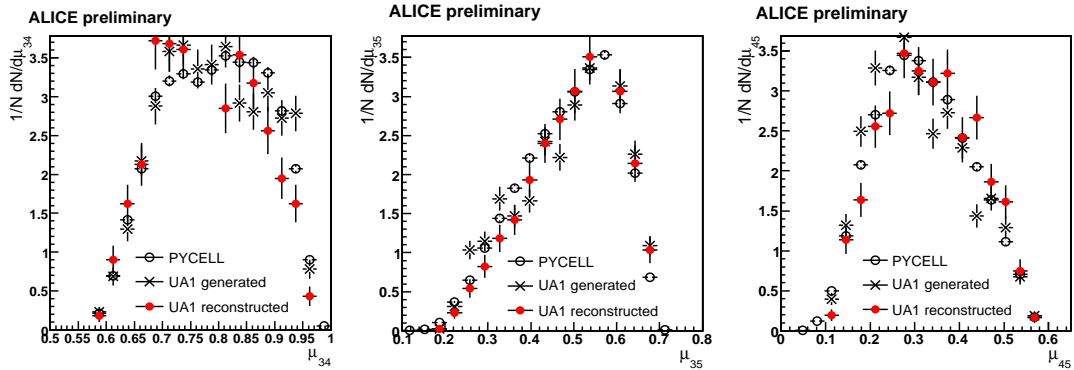}
\caption{Scaled mass distribution for pairs $34$, $35$, $45$ in the selected three jet events. Open circles stand for PYTHIA jets, crosses for generated jets and full circles for reconstructed jets. See text for details.}
\label{fig:ALICEScaled}
\end{figure}

\begin{figure}[ht]
\centering
\includegraphics[height=90mm,width=100mm]{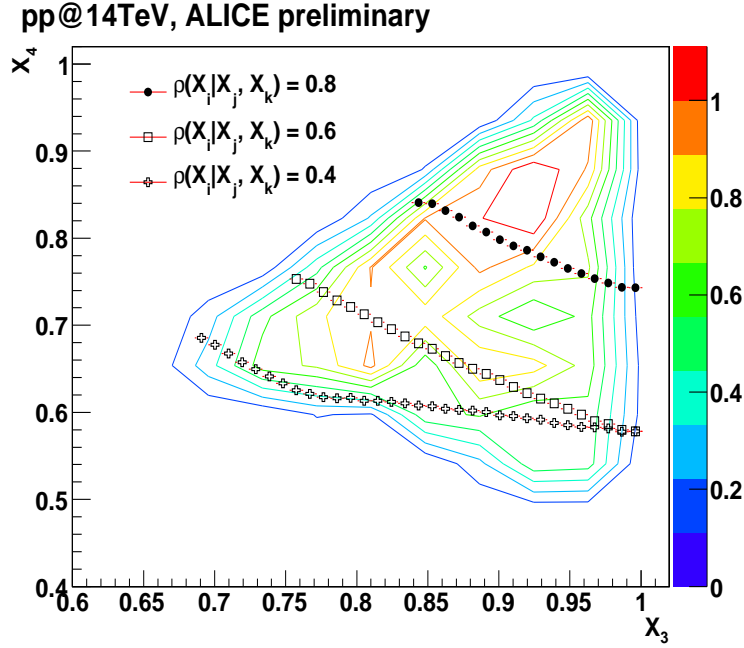}
\caption{Probability that a jet is a gluon distributed over the reconstructed Dalitz plane (contours) for PYTHIA jets. Empty crosses correspond to probability 0.4, empty rectangles to probability 0.6 and full circles to probability 0.8. See text for details.}
\label{fig:ProbGluon}
\end{figure}

In Fig.~\ref{fig:ALICEDalitz} the Dalitz planes for generated (left) and reconstructed (right) three-jet events are shown. In the generated plot there is a clear abundance of events in the upper right corner of the diagram. This means that mostly events with two leading back-to-back jets with a perpendicular low energy jet were found by the jet finder (Fig.~\ref{fig:PYTHIADalitz}). Results are consistent with CDF data.
Reconstructed events show similar behaviour.
\paragraph{}
The scaled mass distributions for different Dalitz variable pairs are shown in Fig.~\ref{fig:ALICEScaled}. Each of the plots displays spectra for PYTHIA (open circles), generated (crosses) and reconstructed (full circles) events and we find good agreement between them for all jet pairs. Furthermore, the shapes and position of peaks are consistent with D\O~ data \cite{Acosta:2004yy}. 
\paragraph{}
The aim of our analysis is to investigate quark and gluon jet properties and for that, proper determination of quark and gluon jet candidates is crucial. What we would like to approach is the possibility to make such a selection based on the shape of the event using Dalitz variables.
\paragraph{}The method we are going to discuss shortly is a probabilistic approach. Using this method we calculate the probability, that a jet carrying energy fraction $X_{i}$ is a gluon, given other two carry energy fractions $X_{j}, X_{k}$ (for details see Ref.~\cite{Fodor:1991az}).
\begin{eqnarray}
\label{eq:ProbGluon}
\rho\left(X_{i}|X{j},X_{k}\right)&=&\frac{X_{j}^{2}+X_{k}^{2}}{\left(1-X_{j}\right)\left(1-X_{k}\right)}\\\nonumber
&\times& \left(\frac{X_{i}^{2}+X_{j}^{2}}{\left(1-X_{i}\right)\left(1-X_{j}\right)}+\frac{X_{j}^{2}+X_{k}^{2}}{\left(1-X_{j}\right)\left(1-X_{k}\right)}+\frac{X_{i}^{2}+X_{k}^{2}}{\left(1-X_{i}\right)\left(1-X_{k}\right)}\right)^{-1}
\end{eqnarray}
The probability is calculated using Dalitz variables and as such can be used to determine gluon jet candidates based on the topology of the event (see Eq.~\ref{eq:ProbGluon}). In Fig.~\ref{fig:ProbGluon}, probabilities of jets being gluons are shown in the Dalitz plane. We do not distinguish whether a jet is quark or gluon initiated, the probability is calculated for each jet from the sample. Only PYTHIA jets were used. Going towards the upper right corner of the plane, probability that a jet is a gluon rises. This means, that this area of the diagram should be mostly populated by gluon jets. To prove this assumption, PYTHIA processes are selected from the sample. In PYTHIA, three-jet events are produced adding initial and final state radiation-a gluon-to the $2\rightarrow 2$ processes (see Table~\ref{tab:PythiaProc}) \cite{PYTHIA:MANUAL}. Most of the jets found are produced in the pure gluonic; $gg\rightarrow gg(+g)$ (process 68), and a mixed channel; $q_{i}g\rightarrow q_{i}g(+g)$ (process 28). This means that our sample is gluon dominated. Combining this fact with an abundance in the vicinity of the upper right corner in reconstructed Dalitz plane (contours in Fig.~\ref{fig:ProbGluon} and Fig.~\ref{fig:ALICEDalitz}-right), we can assume that the most gluons are indeed produced in area suggested by the probability distribution (markers in Fig.~\ref{fig:ProbGluon}).   
%
\section{Summary and conclusions}
The topology of selected three-jet events and the possibility to use it in determining gluon jet candidates was discussed. The results on topology are presented in Figures \ref{fig:ALICEDalitz} and \ref{fig:ALICEScaled}, which show Dalitz plane and scaled masses distributions, respectively. The shapes of the spectra are consistent with CDF and D\O~ results. The dominant shape of selected events corresponds to two back-to-back leading jets with a low energy jet perpendicular with respect to them (see Fig.~\ref{fig:PYTHIADalitz}). 
\paragraph{} To determine the gluon jet candidates, a probabilistic approach was used. According to this approach, most of the gluons in the sample are produced in the vicinity of $[{\textrm X_{3}}, {\textrm X_{4}}]=[1,1]$ corner of the Dalitz plane.
\paragraph{} The discussion presented in these proceedings is based on preliminary results of the three-jet analysis and represents the preparation for real data analysis at ALICE.   

\section*{Acknowledgements}
I would like to thank P\' eter L\' evai, Levente Moln\' ar, Gergely G. Barnaf\" oldi and Andr\' as G. Ag\' ocs for priceless conversations and discussions on presented topic. This work was funded by OTKA grant NK-$062044$ and the NKTH-OTKA grant H$07$-C $70464$.

\end{document}